\documentclass[aps,prd,nofootinbib]{revtex4}

\usepackage{graphicx,bm} \usepackage{amsmath,amssymb}
\usepackage{mathrsfs,verbatim,subfigure}

\begin{document}

\title{Sum rule analysis of vector and axial-vector spectral functions with excited
states in vacuum}

\author{Paul~M.~Hohler}
\email{pmhohler@comp.tamu.edu}
\affiliation{Cyclotron Institute and Department of Physics and Astronomy, Texas A\&M University, College Station, TX 77843-3366, USA}

\author{Ralf~Rapp}
\email{rapp@comp.tamu.edu}
\affiliation{Cyclotron Institute and Department of Physics and Astronomy, Texas A\&M University, College Station, TX 77843-3366, USA}

\begin{abstract}
We simultaneously analyze vector and axial-vector spectral functions in
vacuum using hadronic models constrained by experimental data and the
requirement that Weinberg-type sum rules are satisfied.
Upon explicit inclusion of an excited vector state, {\it viz.}
$\rho^\prime$, and the requirement that the perturbative continua
are degenerate in vector and axial-vector channels, we deduce the
existence of an excited axial-vector resonance state,  $a_1^\prime$,
in order that the Weinberg sum rules are satisfied. The resulting
spectral functions are further tested with QCD sum rules.
\end{abstract}

\maketitle

\section{Introduction}
\label{sec:intro}

Chiral symmetry is believed to be a fundamental symmetry of QCD. However,
this symmetry is spontaneously broken at low temperatures and chemical
potentials by a finite expectation value of the quark (or chiral)
condensate, $\langle 0|\bar{q}q|0 \rangle \simeq -2\, {\rm fm}^{-3}$ per
light quark flavor in vacuum. At higher temperatures, the symmetry is
expected to be restored as the value of the condensate approaches zero.
Such behavior has been observed in lattice-QCD
computations~\cite{Borsanyi:2010bp,Bazavov:2011nk}, but it remains a long-standing
goal to observe chiral restoration in experiment. This is difficult
because a direct measurement of the chiral condensate is not possible.

One way to indirectly infer the chiral condensate and deduce chiral
symmetry restoration is through the use of sum rules.
They are formulated to relate hadronic spectral functions to properties of
the ground state (condensates) including chiral order parameters. The
observation of in-medium changes of hadronic spectral functions can then
signal pertinent changes in these order parameters. Of particular
interest are spectral functions of so-called chiral partners, {\it i.e.},
hadronic states which are degenerate at chiral restoration, but are split
in vacuum due to dynamical chiral breaking. The prime example are the
iso-vector vector and axial-vector channels which are connected through
Weinberg-type sum rules~\cite{Weinberg:1967kj,Das:1967ek,Kapusta:1993hq}
(see, {\it e.g,}, Ref.~\cite{Hilger:2011cq} for a recent study of chiral
partners in the open-charm sector).
Therefore an accurate measurement of both the vector and axial-vector
spectral functions can be used to infer chiral order parameters and thus
``observe" chiral symmetry restoration. Additional information can be
gleaned from the spectral functions by considering QCD sum
rules~\cite{Shifman:1978bx,Shifman:1978by} which are dependent on both
chirally symmetric and chirally breaking ground-state properties. A lot
has been learned about the in-medium vector spectral function from
dilepton measurements and their interpretation~\cite{Rapp:2009yu,Tserruya:2009zt},
but no experimental measurement of the in-medium axial-vector spectral
function has been performed to date. One is thus left with constructing
effective models of the axial-vector spectral function in order to study
chiral symmetry restoration.

In order to reliably evaluate in-medium effects on the vector and
axial-vector spectral functions, one should first control their
properties in vacuum. This is aided by accurate experimental measurements
of the vacuum spectral functions through $\tau$-decays by the ALEPH and
OPAL Collaborations~\cite{Barate:1998uf,Ackerstaff:1998yj}. On the one
hand, sum rules have been used along with experimental data to calculate
the condensates~\cite{Narison:2000cj,Dominguez:2003dr,Bordes:2005wv}.
On the other hand, the condensates can be used to constrain the spectral
functions. Weinberg's original study assumed a $\delta$-function approximation
for both the low-lying vector and axial-vector resonances in his sum rules
to determine the famous relation between the masses of the $\rho$ and $a_1$
mesons, $m_{a_1} = \sqrt{2} m_\rho$~\cite{Weinberg:1967kj}. Similarly, QCD
sum rules have been applied in the vector channel, where early works
approximated the $\rho$ by a $\delta$ function while also considering medium
effects~\cite{Hatsuda:1991ez,Leinweber:1995fn}. More recent work has
considered spectral functions with a Breit-Wigner shape for the $\rho$
peak~\cite{Leupold:1997dg}, and have included the axial-vector channel in
their analyses~\cite{Leupold:2001hj,Kwon:2008vq,Kwon:2010fw}, while others
analyzed calculations based upon a microscopic
theory~\cite{Asakawa:1993pm,Klingl:1997kf,Kwon:2008vq,Kwon:2010fw}.
Clearly, the focus in previous work was on QCD sum rules, with the contribution
to the continuum approximated by a $\theta$-function with a threshold energy.
Furthermore, until the most recent studies~\cite{Kwon:2008vq,Kwon:2010fw}, the
vacuum spectral functions did not take advantage of the high-precision
$\tau$-decay data. The often simplified constructions of spectral
functions used to analyze the sum rules, and the dearth of analyses
to include the axial-vector channel or consider the Weinberg-type sum
rules, leaves this area open for further considerations.

%The vacuum vector spectral function has been extensively studied in
%the literature, while the axial-vector spectral function has been
%studied to a lesser extent. Much of this work have focused on
%developing theories around the microscopic interactions
%Though they have been many such models out there, not many have
%been verified by an analysis of the sum rules.

In the present paper, we simultaneously analyze vector and axial-vector
spectral functions in vacuum using an extended model which combines a
microscopic $\rho$ spectral function with Breit-Wigner ans\"{a}tze for the
$a_1$ and the first excited states. This model is quantitatively constrained
by both the experimental $\tau$-decays~\cite{Barate:1998uf,Ackerstaff:1998yj}
and the Weinberg-type sum rules~\cite{Weinberg:1967kj,Das:1967ek,Kapusta:1993hq}.
By using this
combination of criteria, the model may be considered a non-trivial fit
of the data. Novel features of our analysis include the study of excited
states and the postulate that the continuum contribution is identical
for both the vector and axial-vector channels, as should be the case in
the perturbative regime. In particular, the use of the Weinberg-type sum
rules leads us to deduce the presence of an excited axial-vector state,
about which rather little is known to date~\cite{pdg}. As an additional
check, we utilize the constructed spectral functions in a pertinent analysis
of QCD sum rules.

The outline of this paper is as follows. In Sec.~\ref{sec:sum}, the sum
rules used in our investigation are introduced. In Sec.~\ref{sec:model},
we detail the main ingredients to the vector and axial-vector spectral functions.
Section~\ref{sec:fit} presents the results of our fitting procedure and
discusses our main findings in the context of Weinberg sum rules. In
Sec.~\ref{sec:qcd}, the constructed spectral functions are implemented
into QCD sum rules, and we conclude in Sec.~\ref{sec:conc}.

%%%%%%%%%%%%%%%%%%%%%%%%%%%%%%%%
\section{Weinberg and QCD Sum Rules}
\label{sec:sum}
%%%%%%%%%%%%%%%%%%%%%%%%%%%%%%%%
Sum rules are a valuable tool for understanding non-perturbative
aspects of QCD as they relate two different formulations of the
same correlation function to each other. For the current paper, we
will focus on two classes of sum rules, Weinberg-type and
QCD sum rules.

Let us begin by defining the current-current correlators for the
vector and axial-vector channels,
\begin{equation}
\Pi_V^{\mu\nu}(q) =  - i \int \!d^4x \,e^{i q x} \langle T j_V^\mu(x) j_V^\nu(0)\rangle,
\end{equation}
\begin{equation}
\Pi_A^{\mu\nu}(q) = -i \int \!d^4x\, e^{i q x} \langle T j_A^\mu(x) j_A^\nu(0)\rangle,
\end{equation}
where
$j_V^\mu=\frac{1}{2}\left(\bar{u}\gamma^\mu u -\bar{d} \gamma^\mu d\right)$
and
$j_A^\mu=\frac{1}{2} \left(\bar{u}\gamma^\mu \gamma_5 u-\bar{d} \gamma^\mu \gamma_5 d \right)$
are the pertinent currents in the quark basis. The correlators can be decomposed
into 4-dimensional transverse and longitudinal parts as
\begin{equation}
\Pi_{V,A}^{\mu\nu} (q^2) = \left(-g^{\mu\nu} + \frac{q^\mu q^\nu}{q^2}\right)
\Pi_{V,A}^T(q^2) + \frac{q^\mu q^\nu}{q^2} \Pi_{V,A}^L (q^2) \ .
\end{equation}
%\begin{equation}
%\Pi_A^{\mu\nu}(q^2) = \left(-g^{\mu\nu} + \frac{q^\mu q^\nu}{q^2}\right)
%\Pi_A^T(q^2) + \frac{q^\mu q^\nu}{q^2} \Pi_A^L (q^2) \ .
%\end{equation}
Since the vector current is conserved, $\Pi_V^L = 0$. The longitudinal
component of the axial-vector is governed by the contribution from the
pion which causes the axial current not to be conserved. In vacuum, the
self-energy of the pion is assumed to be negligible, rendering the imaginary
part of the longitudinal axial-vector polarization of the simple form
\begin{equation} \label{eq:pipole}
{\rm Im} \Pi_A^L(s) = - \pi f_\pi^2 s \delta\left(s-m_\pi^2\right)
\end{equation}
with $s=q^2$.

It is also useful to define an additional polarization function,
$\bar{\Pi}_A(q^2) \equiv \Pi_A^T(q^2) + \Pi_A^L(q^2)$ (which corresponds to
$\Pi_A^2$ in Ref.~\cite{Shifman:1978by}).
The transverse polarization functions for both vector and axial-vector
channels are used to define the spectral functions such that
\begin{eqnarray}
\rho_V(q^2) &\equiv -\frac{1}{\pi} {\rm Im} \Pi_V^T(q^2), \\
\label{eq:specdef}
\rho_A(q^2) &\equiv -\frac{1}{\pi} {\rm Im} \Pi_A^T(q^2) \ .
\end{eqnarray}
A spectral function for $\bar{\Pi}_A$ can be defined in a similar manner as
\begin{equation}
\bar{\rho}_A(q^2) \equiv -\frac{1}{\pi} {\rm Im} \bar{\Pi}_A(q^2).
\end{equation}
From the definition of $\bar{\Pi}_A$ and Eqs.~(\ref{eq:pipole}) and
(\ref{eq:specdef}), we see that
$\bar{\rho}_A (s) = \rho_A(s) + f_\pi^2 s \delta\left(s-m_\pi^2\right)$,
{\it i.e.}, this spectral function has contributions from both the pion and
the axial-vector mesons (plus continuum, see below).

The Weinberg-type sum rules characterize moments of the difference
between the vector and axial-vector spectral functions, thereby
quantifying the effects of chiral symmetry breaking. They have the
general form
\begin{equation}
\label{eq:WSR}
\int_0^\infty \! ds\, s^n \Delta\rho(s) = f_n \ ,
\end{equation}
where $\Delta\rho \equiv \rho_V-\rho_A$, $n$ is an integer, and $f_n$
are chiral order parameters dependent on the sum rule in question.

The first two of these sum rules were originally derived by
Weinberg~\cite{Weinberg:1967kj} (hence the name for the class of sum
rules) using current algebra arguments. They read
\begin{eqnarray}
&({\rm WSR}\, 1)& \quad  \int_0^\infty \!ds \,
\frac{\Delta\rho(s)}{s} = f_\pi^2 \ ,
\label{eq:WSR1} \\
&({\rm WSR}\, 2)& \quad \int_0^\infty\! ds\, \Delta\rho(s)
=   f_\pi^2 m_\pi^2 = -2 m_q \langle \bar{q}q \rangle \ ,
\label{eq:WSR2}
\end{eqnarray}
and correspond to $n=-1$ and $n=0$ in Eq.~(\ref{eq:WSR}). Here,
$m_q \simeq 5$ MeV refers to the average current light-quark mass,
while the Gellmann-Oaks-Renner relation~\cite{GellMann:1968rz},
\begin{equation}
 f_\pi^2 m_\pi^2 = -2 m_q \langle \bar{q}q \rangle,
\end{equation}
was used to obtain the second equality in Eq.~(\ref{eq:WSR2}). In
Weinberg's original work, the chiral limit was considered; here we have
included terms linear in the quark
mass~\cite{Pascual:1981jr,Narison:1981ra,Peccei:1986hw,Dmitrasinovic:2000ei}.
Corrections to higher powers in
$m_q$~\cite{Floratos:1978jb,Pascual:1981jr,Narison:1981ra,Peccei:1986hw}
are expected to be small.\footnote{Note that the sum rule in
Ref.~\cite{Floratos:1978jb} labeled as Weinberg's $2^{nd}$ sum rule is not the
same as the one considered here; the one considered here corresponds to the
convergent sum rule in the equal quark-mass limit of Ref.~\cite{Floratos:1978jb}.}
One sees that the chiral condensate can be calculated using
Eq.~(\ref{eq:WSR2}) should both the vector and axial-vector spectral
functions be known precisely. The sum rule for $n=-2$ was introduced by
Das, Mathur, and Okubo \cite{Das:1967ek},
\begin{equation}
\label{eq:WSR0}
({\rm WSR}\, 0) \quad \quad \int_0^\infty \!ds\,
\frac{\Delta\rho(s)}{s^2} =
\frac{1}{3} f_\pi^2 \langle r_\pi^2\rangle -F_A \ ,
\end{equation}
where $\langle r_\pi^2 \rangle$ is the mean squared radius of the charged
pion, and $F_A$ is the coupling constant for the radiative pion decay,
$\pi^\pm \rightarrow \mu^\pm \nu_\mu \gamma$. We label this sum rule as
the $0^{\rm th}$ one since it has a smaller value of $n$ as compared
to Weinberg's initial sum rules. Lastly, Kapusta and
Shuryak~\cite{Kapusta:1993hq} derived a sum rule for
$n=1$,
\begin{equation}
\label{eq:WSR3}
({\rm WSR}\, 3) \quad \quad \int_0^\infty ds s \Delta\rho(s)
= - 2 \pi \alpha_s \langle \mathcal{O}_4 \rangle \ ,
\end{equation}
where $\langle \mathcal{O}_4 \rangle$ is the part of the four-quark condensate
which breaks chiral symmetry. The explicit quark content of this operator is
given by
\begin{equation} \label{eq:Q4XB}
\langle \mathcal{O}_4 \rangle = \frac{1}{4} \left(\left\langle \left(\bar{u}\gamma_\mu\gamma_5\lambda^a u-\bar{d}\gamma_\mu\gamma_5\lambda^ad\right)^2 - \left(\bar{u}\gamma_\mu\lambda^au-\bar{d}\gamma_\mu\lambda^ad\right)^2\right\rangle\right),
\end{equation}
where $\lambda^a$ denote the Gell-Mann matrices. It is common to assume that this
operator can be factorized into the chiral condensate,
\begin{equation}
\label{eq:factor}
\langle \mathcal{O}_4 \rangle = \frac{16}{9} \kappa
\langle \bar{q}q \rangle^2 \ ,
\end{equation}
where $\kappa$ is a parameter larger than one to mimic the contributions beyond
ground-state saturation. Typically the Weinberg-type sum rules are expressed in
vacuum in terms of $\rho_A$ as is done here. However, it is possible to include
the pion pole into the axial-vector spectral function and thereby write the sum
rules in terms of $\bar{\rho_A}$ with their right-hand-side (RHS) appropriately
adjusted. This was done in Ref.~\cite{Kapusta:1993hq} for the sum rules at finite
temperature.

For studies of chiral symmetry restoration, satisfying these sum rules is
critical both in medium and in vacuum. In the present work, we will make
the first step by using the first three,
Eqs.~(\ref{eq:WSR1}), (\ref{eq:WSR2}), and (\ref{eq:WSR0}), to constrain
the parameters of the vacuum spectral functions. The last sum rule,
Eq.~(\ref{eq:WSR3}), is not used due to a large uncertainty in the value
of $\kappa$ and the rather high sensitivity of the integration to large $s$
where the control from experimental data is limited. Nevertheless,
we will still examine how well it is satisfied by the constructed spectral
functions.

The QCD sum rules, on the other hand, apply to the vector and axial-vector
channels separately. They were first introduced by Shifman, Vainshtein
and Zakharov using the operator product expansion
(OPE)~\cite{Shifman:1978bx,Shifman:1978by}.
They relate the integral over the spectral function to a series of
ground-state operators (expectation values) specific for the channel of
interest. To improve the convergence of the integral over the spectral
function, one performs a Borel transform on both sides of the sum rule.
For the vector channel one obtains
\begin{equation}
\label{eq:rhoQCD}
\frac{1}{M^2}\!\int_0^\infty \!ds \frac{\rho_V(s)}{s} e^{-s/M^2}
= \frac{1}{8\pi^2} \left(1+\frac{\alpha_s}{\pi}\right)
+\frac{m_q \langle\bar{q}q\rangle}{M^4}
+\frac{1}{24 M^4}\langle\frac{\alpha_s}{\pi} G_{\mu\nu}^2\rangle
- \frac{56 \pi \alpha_s}{81 M^6}  \langle \mathcal{O}_4^V \rangle \ldots \, ,
\end{equation}
and for the axial-vector channel
\begin{equation}
\label{eq:a1QCD}
\frac{1}{M^2}\!\int_0^\infty \!ds \frac{\bar{\rho}_A(s)}{s} e^{-s/M^2}
= \frac{1}{8\pi^2} \left(1+\frac{\alpha_s}{\pi}\right)
+\frac{m_q \langle\bar{q}q\rangle}{M^4}
+\frac{1}{24 M^4}\langle\frac{\alpha_s}{\pi} G_{\mu\nu}^2\rangle
+ \frac{88 \pi \alpha_s}{81 M^6}  \langle \mathcal{O}_4^A \rangle \ldots \, .
\end{equation}
%The same set of sum rules can be found in Ref.~\cite{Leupold:2001hj}
%with the replacement $\rho(s)/s = {\rm Im} R(s)$.
%These sum rules were first constructed in \cite{Shifman:1978bx,Shifman:1978by}.
By performing the Borel transform, one trades the space-like 4-momentum,
$q^2 = -Q^2$, with the Borel mass, $M^2$. Note that the axial-vector spectral
function is defined to contain the contribution from the
pion pole (use of $\bar{\rho}_A$, not $\rho_A$). To the order $1/M^6$,
which we are working, the following operators figure: the chiral condensate,
$\langle \bar{q}q \rangle$, the gluon condensate,
$\langle \frac{\alpha_s}{\pi} G^2_{\mu\nu} \rangle$, and the vector and
axial-vector four-quark condensates, $\langle \mathcal{O}_4^V \rangle$ and
$\langle \mathcal{O}_4^A \rangle$, respectively. The four-quark condensates can
be expressed in terms of their quark content as in \cite{Leupold:2001hj}
\begin{equation}
\langle \mathcal{O}_4^V \rangle = \frac{81}{224}\left\langle \left(\bar{u}\gamma_\mu\gamma_5\lambda^au-\bar{d}\gamma_\mu\gamma_5\lambda^ad\right)^2\right\rangle + \frac{9}{112}\left\langle\left(\bar{u} \gamma_\mu\lambda^a u+\bar{d}\gamma_\mu\lambda^a d\right) \sum_{q = u,d,s} \bar{q} \gamma^\mu\lambda^a q\right\rangle,
\end{equation}
\begin{equation}
\langle \mathcal{O}_4^A \rangle = -\frac{81}{352}\left\langle \left(\bar{u}\gamma_\mu\lambda^au-\bar{d}\gamma_\mu\lambda^ad\right)^2\right\rangle - \frac{9}{176}\left\langle\left(\bar{u} \gamma_\mu\lambda^a u+\bar{d}\gamma_\mu\lambda^a d\right) \sum_{q = u,d,s} \bar{q} \gamma^\mu\lambda^a q\right\rangle,
\end{equation}
and are related to the chirally breaking four-quark condensate of Eq.~(\ref{eq:Q4XB}) as
\begin{equation}
\langle \mathcal{O}_4 \rangle = \frac{16}{9}\left(\frac{7}{18} \langle \mathcal{O}_4^V \rangle + \frac{11}{18} \langle \mathcal{O}_4^A \rangle \right).
\end{equation}
These forms of the four-quark condensates and the coefficients in the sum rules are chosen such that $\langle \mathcal{O}_4^V \rangle$ and $\langle \mathcal{O}_4^A \rangle$ have a simple factorized form, {\it viz.} $\langle \mathcal{O}_4^V \rangle
= \kappa_V \langle \bar{q}q \rangle^2$ and $\langle \mathcal{O}_4^A \rangle
= \kappa_A \langle \bar{q}q \rangle^2$. The parameters
$\kappa_V$ and $\kappa_A$ are, in principle, independent and thus could take on
different values; they are related to $\kappa$ in Eq.~(\ref{eq:factor}) via
\begin{equation}
\label{eq:kappas}
\kappa = \frac{7}{18} \kappa_V + \frac{11}{18} \kappa_A \ .
\end{equation}
For simplicity we have chosen $\kappa_V$ and $\kappa_A$ as numerically identical,
which then implies from Eq.~(\ref{eq:kappas}) that $\kappa$ is also numerically
the same, {\it viz.} $\kappa_V = \kappa_A = \kappa$. A more general discussion on
the properties of four-quark condensates can be found in \cite{Thomas:2007gx}.
We note that the sign
of the chiral-condensate term is the same for the two sum rules, not opposite
as one might expect since the chiral condensate is a chirally
odd operator. The latter reasoning actually applies to the transverse part of
the axial-vector current. However, the QCD sum rule written above is for
the total current including the longitudinal part induced by the pion
contribution. This leads to the signs presented in Eq.~(\ref{eq:a1QCD}).
We also note that the Weinberg sum rules 1-3 can be derived from the QCD
sum rules by subtracting both sides of the vector and axial-vector QCD
sum rules from each other, Taylor expanding the Borel convergence factor
$e^{-s/M^2}$, and equating the coefficients of equal powers of $M^2$ on
each side. A similar procedure was pointed out in
Ref.~\cite{Shifman:1978by} and used in Ref.~\cite{Kapusta:1993hq} to
derive Eq.~(\ref{eq:WSR3}).

If one knows the vector or axial-vector spectral function, the QCD sum
rule can be used to determine the values of the condensates. Conversely,
since the condensates are universal low-energy operators, they can be
calculated from other processes, and in turn be used to constrain the
spectral functions. Though related to the Weinberg-type, QCD sum rules
can provide different constraints on the spectral functions because both
chirally even and chirally odd operators are involved. A recent
study of the constraints on the spectral functions imposed by the chiral
odd operators in the QCD sum rules can be found in Ref.~\cite{Hilger:2010cn}.
A potential drawback of the QCD sum rules comes from the relatively large
uncertainty of the values of the gluon and 4-quark condensates, as compared
with the pion mass or its decay constant in the first three Weinberg sum rules.
Therefore, we will not use the QCD sum rules to constrain the vacuum
spectral functions, but rather use the constructed spectral functions to
constrain the condensates and in this way perform a consistency check.

In principle, there are corrections to the QCD sum
rules~\cite{Shifman:1978bx,Shifman:1978by,Novikov:1981xj}. Besides the
conventional perturbative and power corrections, the most prominent
one not included here is associated with the exchange of instantons with
the vacuum. It turns out that the importance of instantons depends
on the specific correlator~\cite{Novikov:1981xj}, {\it i.e.}, whether
direct-instanton interactions are operative, like in the scalar and
pseudoscalar meson channels. In both the vector and axial-vector
channels these are absent and the remaining corrections are
probably small~\cite{Novikov:1981xj}. We therefore believe that
the QCD sum rules considered here are sufficiently accurate.

%%%%%%%%%%%%%%%%%%%%%%%%%%%%%%%%%%%%%%%%%%%%%%%%%%%%%%%%%%%%%%%%%%5
\section{Hadronic Models for Vector and Axial-vector Spectral Functions}
\label{sec:model}
%%%%%%%%%%%%%%%%%%%%%%%%%%%%%%%%%%%%%%%%%%%%%%%%%%%%%%%%%%%%%%%%%%%%
So far, we have discussed Weinberg-type and QCD sum rules and their
usefulness in constructing vacuum spectral functions. In this section,
we present a model which we will use to construct the spectral functions
suitable for a physically motivated fit to $\tau$-decay data. Note that
for the purpose of the present work the concrete fit functions and parameter
values are not important as long as the data are accurately reproduced.

Our main ansatz, which is one of the differences of our work from
previous analyses, is that the vector and axial-vector spectral functions
are divided up into three parts: the ground-state resonance, a first
excited state, and a universal continuum,
\begin{eqnarray}
\rho_V(s) &=& \rho_V^{\rm gs}(s) + \rho_V^{\rm ex}(s) + \rho^{\rm cont}(s)
\\
\rho_A(s) &=& \rho_A^{\rm gs}(s) + \rho_A^{\rm ex}(s) + \rho^{\rm cont}(s).
\end{eqnarray}
The explicit form of each part will be discussed in turn.

The ground-state resonance in the vector channel, the $\rho(770)$,
has been well-studied in effective hadronic Lagrangians; we here employ
the spectral function of Ref.~\cite{Urban:1999im} which was originally
fit to the pion electromagnetic form-factor and $P$-wave $\pi\pi$
scattering phase shifts, but also turns out to
describe the experimental $\tau$-decay data well~\cite{Rapp:2002tw};
its medium modifications have been widely applied to experiment, {\it e.g.},
in dilepton and photon production in heavy-ion collisions~\cite{Rapp:1999us}
and in elementary reactions~\cite{Riek:2008ct}. These features make the
$\rho$ spectral function a suitable starting point (both in vacuum and for
future studies in medium) for the other two components which will
largely rely on fits to the $\tau$-decay data.

The spectral properties of the $a_1$ meson are much less studied,
especially in the medium. Since our eventual objective are studies of
the spectral function's medium modifications (and the pattern of chiral
symmetry restoration for the vector and axial-vector), we adopt a more
schematic ansatz for the $a_1$ spectral function. Rather than using the
vacuum spectral function from an effective hadronic model (see, {\it e.g.},
Refs.~\cite{Kim:1999pb,Urban:2001ru,Harada:2005br,Wagner:2007wy,Struber:2007bm,Wagner:2008gz,Cabrera:2009ep}),
we employ a generic Breit-Wigner form,
\begin{equation}
\label{eq:b-w}
\rho_{a_1} (s) = \frac{1}{\pi} \frac{M_{a_1}^4}{g_{a_1}^2}
\frac{\sqrt{s}\, \Gamma_{a_1}(s)}
{\left(s-M_{a_1}^2\right)^2 + s \Gamma_{a_1}(s)^2} \ ,
\end{equation}
where $M_{a_1}$ is the bare mass, $g_{a_1}$ the axial-vector coupling
constant, and $\Gamma_{a_1}(s)$ the energy dependent width of the resonance.
These parameters will be determined uniquely for the $a_1$. The form of
the coefficient $M_{a_1}^4/g_{a_1}^2$ is the same as in vector-meson
dominance. The energy dependence of the $a_1$ width will be approximated by
an $S$-wave decay into $\rho\pi$. In order to properly incorporate the
three-pion final state, the spectral shape of the $\rho$ will be accounted
for by integrating over its spectral function. Finally,
to simulate the finite size of the $\rho \pi a_1$ vertex and to control
the large energy behavior of the $a_1$ spectral function, a hadronic
form-factor is included. Putting all of this together, the energy dependent
width of the $a_1$ takes the form
\begin{equation}
\Gamma_{a_1}(s) = -\int_0^\infty \! ds'\, \frac{{\rm Im} D_\rho(s')}{\pi}
\Gamma^0_{a_1} \frac{p_{\rm cm}^{\rho \pi}(s,s')}{p_{\rm cm}^{\rho \pi}(M_{a_1}^2,s')}
\left(\frac{ \Lambda_{a_1}^2+M_{a_1}^2}{ \Lambda_{a_1}^2+s}\right)^2 \, ,
\end{equation}
where $D_{\rho}(s)$ is the $\rho$ propagator, and the center of mass momentum of the $\rho$ and $\pi$ decay products, $p_{\rm cm}^{\rho \pi}(s,s')$, is given by
\begin{equation}
p_{\rm cm}^{\rho \pi}(s,s')=\frac{1}{2}\left(\frac{\left(s+s'-M_\pi^2\right)^2-4 s s'}{s}\right)^{1/2}.
\end{equation}
This ansatz involves two additional parameters, a form-factor cut-off, $\Lambda_{a_1}$,
and the $\rho\pi a_1$ coupling constant which determines the magnitude of the on-shell
width, $\Gamma^0_{a_1}$, defined as $\Gamma_{a_1}^0=\Gamma_{a_1}(M_{a_1}^2)$.

The first excited resonances in both vector and axial-vector channels,
namely the $\rho^\prime$ and $a_1^\prime$, are included in the current
construction. Their contribution to the spectral functions is presented
here while the motivation will be provided later. Again, we note that
not only is this the first attempt to include them in such a study,
but given the well established $\rho^\prime$ contribution, we will be
deducing the \emph{need} to include the $a_1^\prime$ based upon the
Weinberg-type sum rules. A microscopic description of the excited
resonance states is not well understood to date. Therefore, as was done
for the $a_1$, the spectral functions of the excited resonances will be
approximated by a Breit-Wigner shape as in Eq.~(\ref{eq:b-w}). To
construct the energy-dependent width, we are once again guided by plausible
decay channels. However, unlike for the $a_1$ case, the decay products
are not well established. Instead we postulate that the width exhibits
a threshold which is controlled by an estimated mass scale. We further
postulate that the partial-wave distribution of the decay products is
the same as for its corresponding ground state, namely the $\rho^\prime$
decays through a $P$-wave process while the $a_1^\prime$ decays through
an $S$-wave process. Furthermore, since the actual decay products of
the excited resonances are not explicitly specified, treating them
off-shell by folding in their spectral function is not warranted.
As for the ground states, a form-factor will be included.
This leads to a width of the $\rho^\prime$ as
\begin{equation}
\Gamma_{\rho^\prime} (s) = \Gamma^0_{\rho^\prime}
\left(\frac{s-\left(M_{th}^{(\rho^\prime)}\right)^{2}}{M_{\rho^\prime}^2 -
\left(M_{th}^{(\rho^\prime)}\right)^{2}}\right)^{3/2}
\frac{M_{\rho^\prime}^2}{s} \left(\frac{ \Lambda_{\rho^\prime}^2
+ M_{\rho^\prime}^2}{ \Lambda_{\rho^\prime}^2 + s}\right)^2 \ ,
\end{equation}
and for the $a_1^\prime$ as
\begin{equation}
\Gamma_{a_1^\prime}(s) = \Gamma^0_{a_1^\prime}
\left(\frac{s-\left(M_{th}^{(a_1^\prime)}\right)^{2}}{M_{a_1^\prime}^2
- \left(M_{th}^{(a_1^\prime)}\right)^{2}}\right)^{1/2}
\left(\frac{ \Lambda_{a_1^\prime}^2
+ M_{a_1^\prime}^2}{ \Lambda_{a_1^\prime}^2 + s}\right)^2 \ .
\end{equation}
The parameters $M_{\rho^\prime}$ and $M_{a_1^\prime}$ set the bare mass
for the pertinent resonances; $M_{th}^{(\rho^\prime)}$ and
$M_{th}^{(a_1^\prime)}$ correspond to the effective mass thresholds of the
decay products, while $\Lambda_{\rho^\prime}$ and $\Lambda_{a_1^\prime}$
set the form-factor scales. Lastly, the overall strengths of the widths are
given by $\Gamma^0_{\rho^\prime}$ and $\Gamma^0_{a_1^\prime}$, which
again have been normalized such that
$\Gamma_{\rho^\prime}(M_{\rho^\prime}^2) = \Gamma_{\rho^\prime}^0$ and
$\Gamma_{a_1^\prime} (M_{a_1^\prime}^2) = \Gamma_{a_1^\prime}^0$.

At high energies, the contribution from the continuum can be calculated from
perturbative QCD. Most previous work has assumed that this contribution
exhibits a threshold energy below which it vanished while the perturbative
value was assumed above. We have adopted a continuous
function for all energies which approaches the perturbative value for large
energies~\cite{Kapusta:1993hq},
\begin{equation}
\rho^{\rm cont}(s) = \frac{s}{8\pi^2} \left(1+\frac{\alpha_s}{\pi}\right)
\left( \frac{1}{1+ \exp [\left(E_{th}-\sqrt{s}\right)/ \delta]}\right).
\end{equation}
The parameter $E_{th}$ plays the role of a threshold energy while
$\delta$ determines how fast the limiting value is achieved.
Furthermore, because at high energies, {\it i.e.}, in the perturbative regime,
QCD is chirally invariant, we postulate an identical continuum contribution
to both channels.
Thus the parameters $E_{th}$ and $\delta$ are the
same for the vector and axial-vector channels and
$\rho_V^{\rm cont} = \rho_A^{\rm cont}$. This is the first time that this
feature has been implemented; besides the underlying physical motivation of
degenerate perturbative continua, it will play a crucial role in interpreting
the vacuum spectral functions.

%%%%%%%%%%%%%%%%%%%%%%%%%%%%%%%%%%%%%%%%%%%%%%%%%%%%%%%%%%%%%%%%%%%%
\section{Vacuum Spectral Functions and Discussion}
\label{sec:fit}
\begin{figure}[htb]
\centering
\subfigure[Vector Spectral Function]{\label{fig:rhospec}
\includegraphics[width=.45\textwidth]{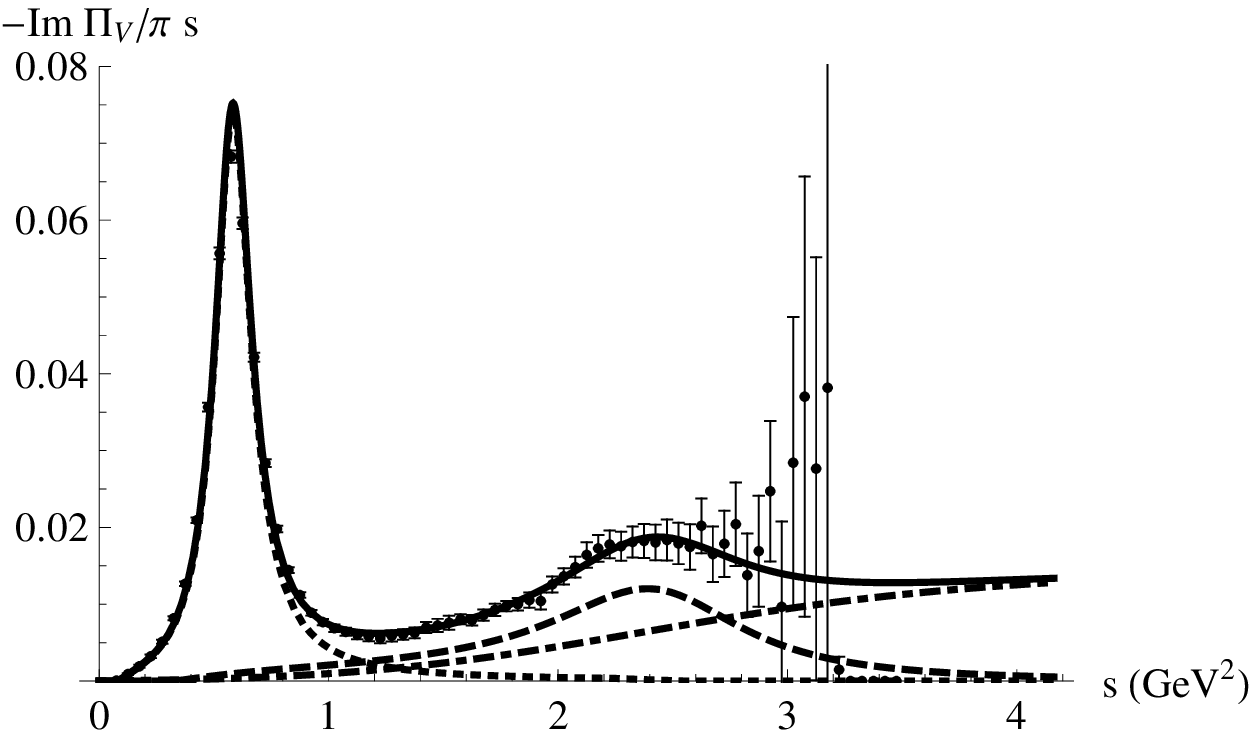}}
\subfigure[Axial-Vector Spectral Function]{\label{fig:a1spec}
\includegraphics[width=.45\textwidth]{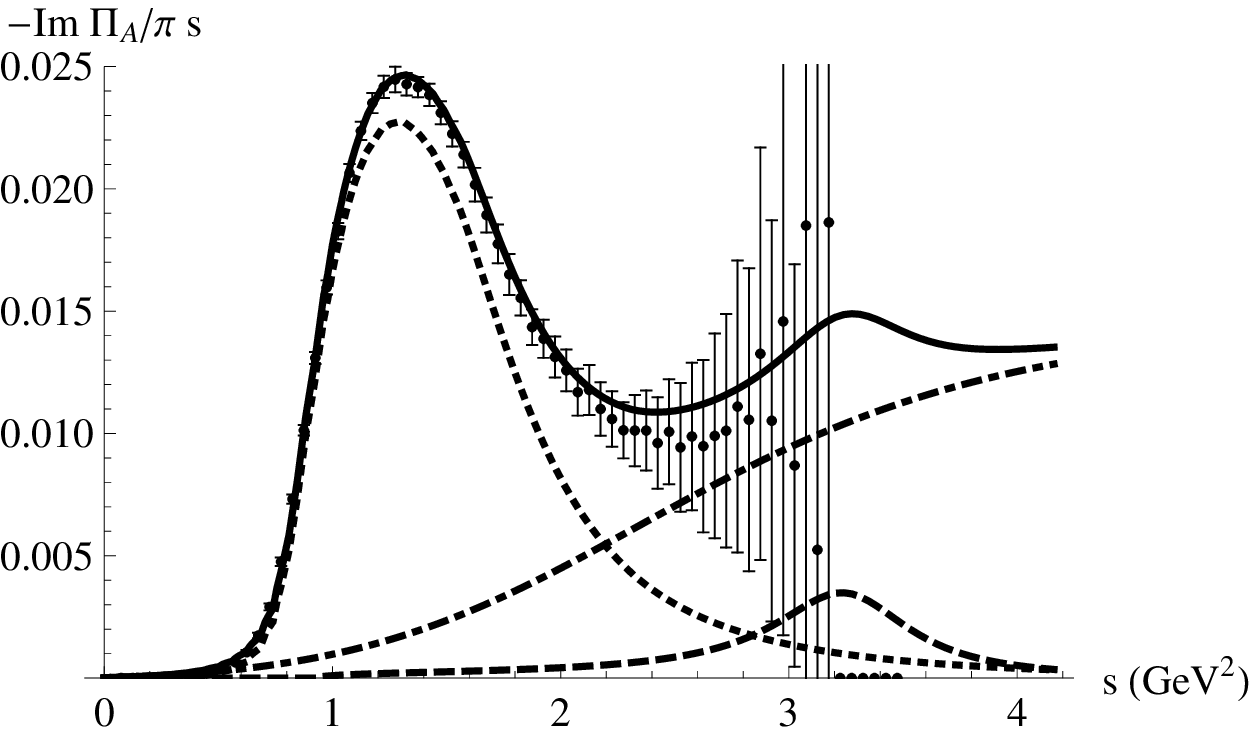}}
\caption{Spectral functions for the vector (left) and axial-vector channel
(right) compared to experimental data for hadronic $\tau$ decays by the
ALEPH collaboration~\cite{Barate:1998uf}. The different curves highlight
the contributions to the total spectral function (solid curve) from the
ground-state resonance (dotted curve), the excited resonance (dashed
curve), and the continuum (dot-dashed curve).}
\label{fig:spec}
\end{figure}

%%%%%%%%%%%%%%%%%%%%%%%%%%%%%%%%%%%%%%%%%%%%%%%%%%%%%%%%%%%%%%%%%%%%
The total number of fit parameters of our model for the vector and axial-vector
spectral functions in vacuum is 16 (the three from the $\rho$ spectral function have been fixed before). They include the masses $M_X$, the
couplings $g_X$, the width strengths $\Gamma^0_X$, and the form-factors
scales $\Lambda_X$ for each $X=a_1, \rho^\prime$ and $a_1^\prime$ states. The
$\rho^\prime$ and $a_1^\prime$ furthermore incorporate a threshold mass
scale, $M_{th}^{(X)}$, and finally the continuum carries the two parameters
$E_{th}$ and $\delta$. The continuum also depends on the strong coupling constant
which was chosen to be $\alpha_s(1 {\rm GeV})=0.5$. This set of convenient
and physically motivated parameters was determined by fitting the spectral
distributions of hadronic $\tau$-decays into an even and odd number of
pions~\cite{Barate:1998uf,Ackerstaff:1998yj}.
In addition, the requirement of reproducing the Weinberg-type sum rules was
imposed (except the third one, Eq.~(\ref{eq:WSR3}), {\it cf.}~our comment above).
The fit was not done for all 16 parameters simultaneously, but rather three
sequential fits were performed. First, the parameters of the $a_1$ peak and the
continuum were determined from the axial-vector ALEPH $\tau$-decay data. Second,
the $\rho'$ peak parameters were determined from the vector ALEPH $\tau$-decay
data. And lastly, the parameters associated with the $a_1^\prime$ peak were
determined from the Weinberg-type sum rules.
The parameter values are summarized in Tab.~\ref{tab:para}.
Figure~\ref{fig:spec} shows the resulting spectral functions, divided by $s$ to
render them dimensionless, compared with the experimental data. A measure of the goodness of fit
is the coefficient of determination, $R^2$. The vector channel
has an $R^2$ value of 0.991 while the axial-vector channel (without the $a_1^\prime$)
has an $R^2$ value of 0.997. Both of these indicate very good agreement of the
resulting spectral functions and the experimental data.

\begin{table}[htb]
\begin{center}
\begin{tabular}{|c|c|c|c|}
\hline
& $a_1$ & $ \rho^\prime$& $a_1^\prime$ \\
\hline
$M_{X}$ &1.246  GeV &1.565 GeV &1.802 GeV\\
$\Gamma^0_{X}$ &0.612  GeV&0.32  GeV &0.2 GeV\\
$g_{X}$ &6.15  &11.44&28.70\\
$\Lambda_{X}$ &0.61  GeV&1.41 GeV& 1.24 GeV\\
$M_{th}^{(X)}$ & N/A (3$\pi$) &0.56 GeV&0.96 GeV\\
\hline
\end{tabular}
\quad \quad
\begin{tabular}{|c|c|}
\hline
\multicolumn{2}{|c|}{Continuum parameters}\\
\hline
$E_{th}$ & 1.60 GeV\\
$\delta$ & 0.227 GeV\\
\hline
\end{tabular}
\quad \quad
\begin{tabular}{|c|c|}
\hline
\multicolumn{2}{|c|}{Constant parameters}\\
\hline
$m_\pi$ & 139.6 MeV\\
$f_\pi$ & 92.4 MeV\\
$m_q$ & 5 MeV\\
$\alpha_s(1 {\rm GeV})$ & 0.5\\
$\langle r^2_\pi \rangle$ & 0.439 ${\rm fm}^2$\\
$F_A$ & 0.0058\\
$\langle \bar{q} q \rangle$ & $(-0.25 {\rm GeV})^3$ \\
\hline
\end{tabular}
\end{center}
\caption{List of parameters of the constructed model as constrained by
experimental data and the Weinberg-type sum rules.}
\label{tab:para}
\end{table}

\begin{figure}[!tb]
\centering \subfigure[WSR 0]{ \label{fig:wsr0}
\includegraphics[width=.45\textwidth]{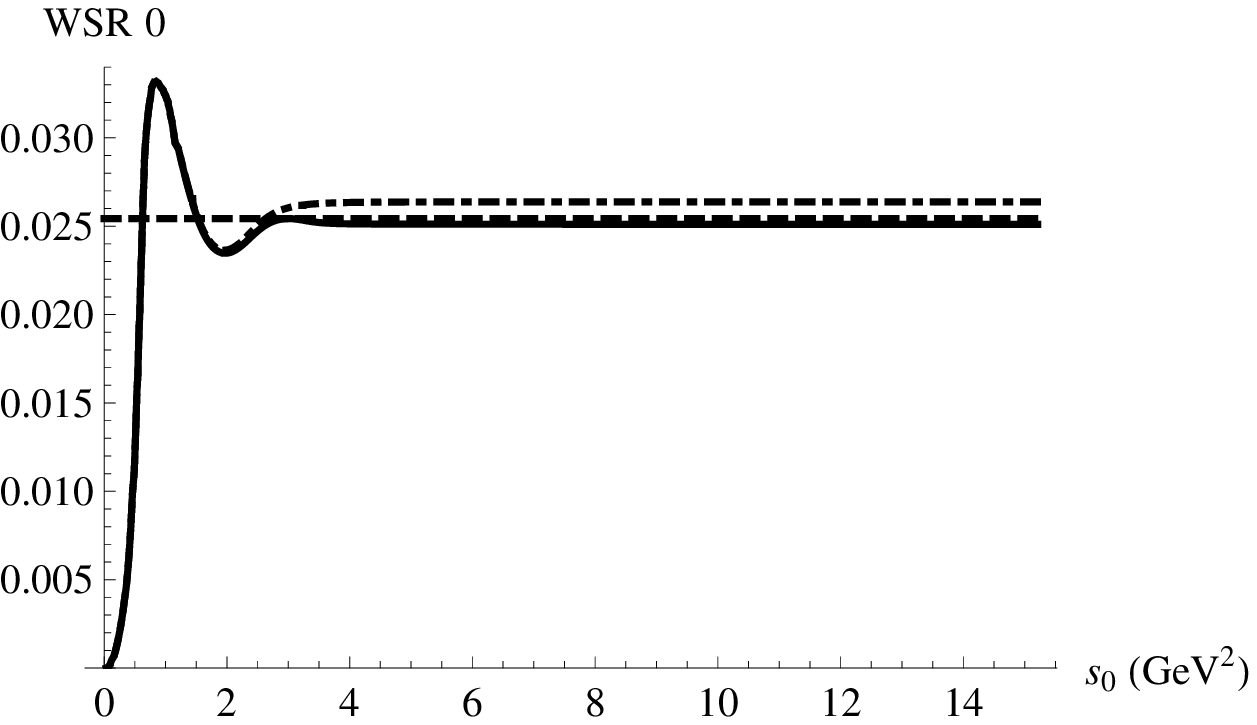}}
\subfigure[WSR 1]{ \label{fig:wsr1}
\includegraphics[width=.45\textwidth]{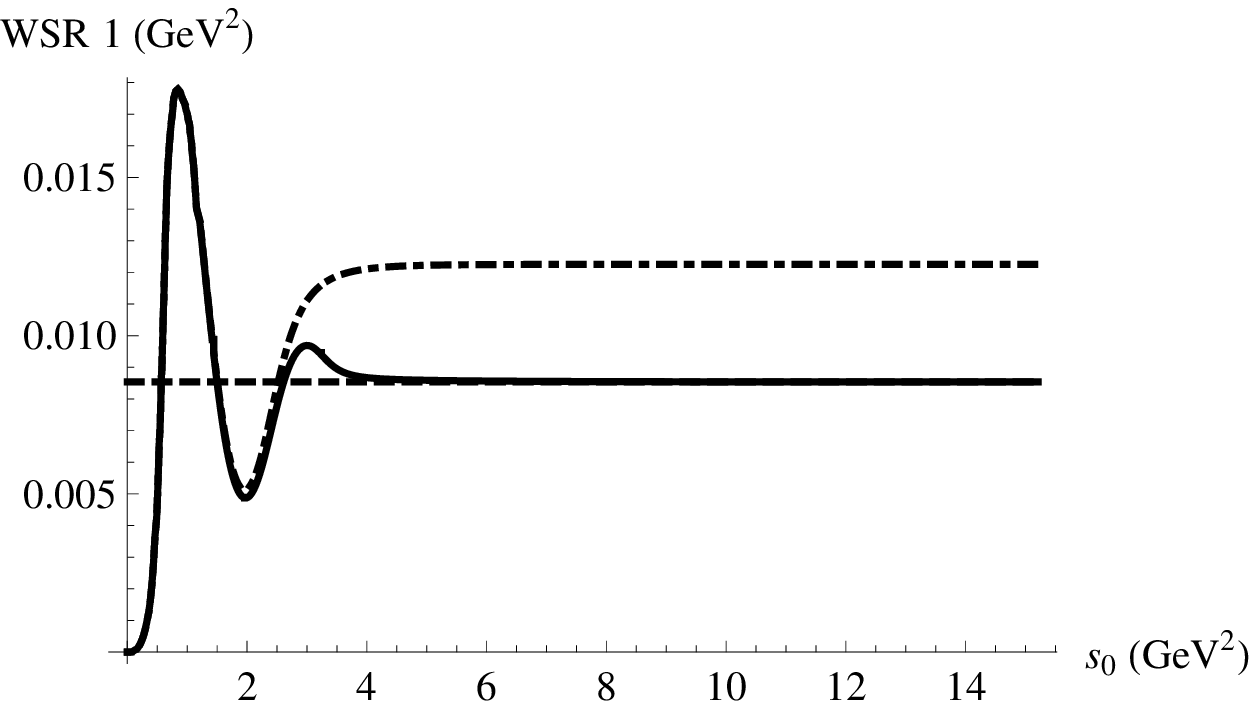}}
\subfigure[WSR 2]{ \label{fig:wsr2}
\includegraphics[width=.45\textwidth]{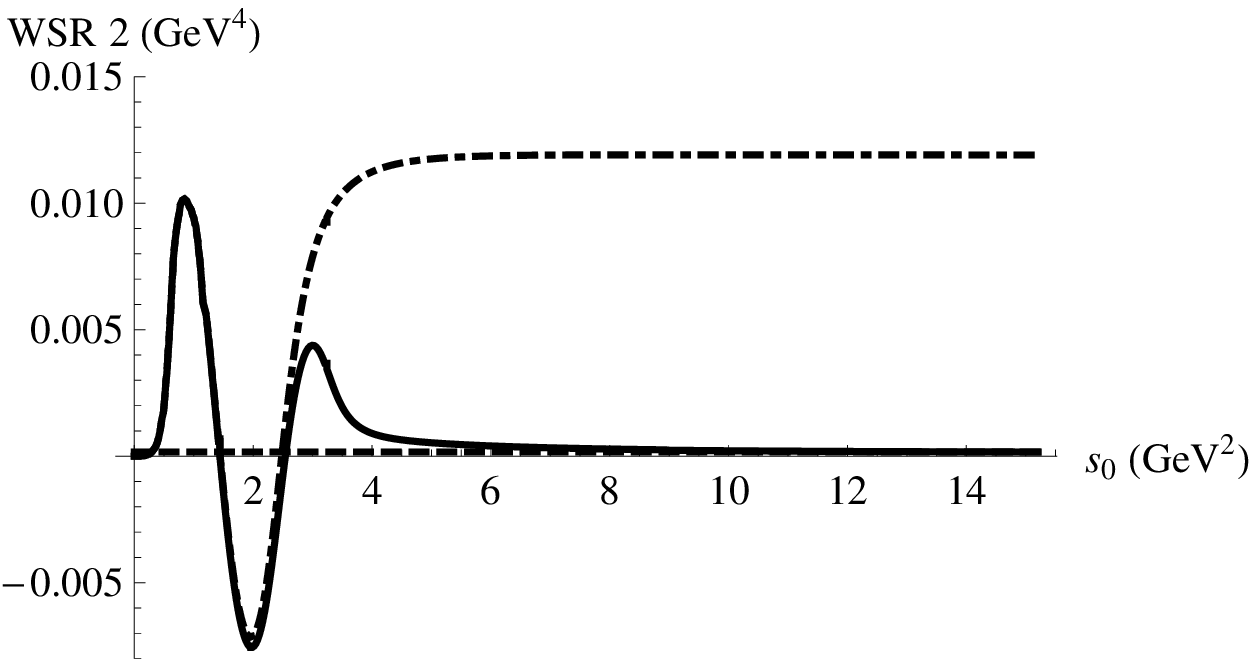}}
\subfigure[WSR 3]{ \label{fig:wsr3}
\includegraphics[width=.45\textwidth]{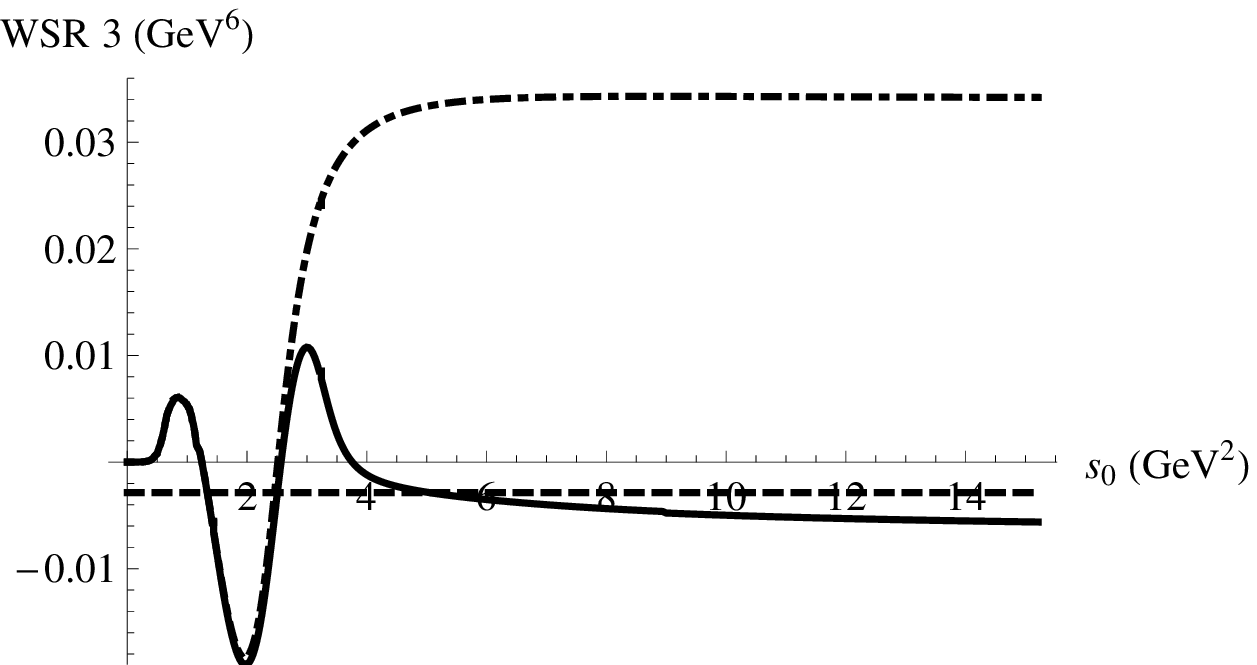}}
\caption{Graphical representation of the Weinberg-type sum rules. The LHS of
Eqs.~(\ref{eq:WSR1}), (\ref{eq:WSR2}), (\ref{eq:WSR0}) and (\ref{eq:WSR3}) is
plotted as a function of the upper integration limit (solid curve), compared
to the theoretical experimental value for the RHS (dashed curve). The
dot-dashed curve is the same as the solid curve but excluding the contribution
from the $a_1^\prime$ state.}
  \label{fig:wsr}
\end{figure}

The results for the Weinberg-type sum rules are graphically assessed in
Fig.~\ref{fig:wsr}. Here the left-hand-side (LHS) of each sum rule is
plotted as a function of the upper limit of the energy integration. Toward
high energies, {\it i.e.}, when the spectral functions degenerate so that
their difference no longer contributes to the sum rule, the curve should
converge to the value of the RHS, represented by the dashed curve.
Table~\ref{tab:WSR} quantifies the numerical deviation of the asymptotic
value from the RHS. The values are quoted such that a positive (negative)
deviation means that the contribution from the vector (axial-vector)
channel is too large. Primarily by introducing, and then adjusting the
mass and the coupling of the $a_1^\prime$ state, ``perfect" numerical
agreement with the RHS of Weinberg-type sum rules 1 and 2 can be achieved.
This was intentionally done since the respective RHS of these two sum rules are
known with much better precision than for WSR-0 and especially WSR-3. Not
surprisingly then, the latter exhibits the largest deviation.
The value of $\kappa$ which we use to quantify this deviation is
actually determined from the QCD sum rules which will be described in the
next section. In all cases, even for WSR-3, the deviation of the converged
value from the asymptotic value dictated by the RHS is small compared to
the size of the oscillations seen at low energies. The WSR-3 is  more
sensitive to the higher energy regime as compared to the other sum rules
because of the larger power of $s$ in the integral in Eq.~(\ref{eq:WSR3}).
Its deviation suggests that a little less spectral strength is needed in
the axial-vector channel at high energies. Nonetheless, the ``excess"
axial-vector strength in WSR-3 is less than 10\% of
the total strength generated by the $a_1^\prime$.  Another
possibility is the introduction of a second excited vector state at higher
energies; we refrain from this possibility due to the lack of constraints
available by the current scheme.

A few comments pertaining to our fit are in order. First, because we have
postulated the continuum for the vector and axial-vector spectral functions to
be identical, the energy of the effective threshold is larger than in previous
applications. This is essentially dictated by the dip region in the
axial-vector spectral function data around $s=2.2 \,{\rm GeV}^2$,
which lies below the pQCD continuum level of
$\frac{1}{8 \pi^2}\left(1+\alpha_s/\pi\right) \simeq 0.015$.
The onset of the continuum must therefore be pushed out to higher energies
to accommodate this dip. This can be observed in Fig.~\ref{fig:a1spec} by
inspecting the contribution from the continuum, the dot-dashed curve. The
need for larger $E_{th}$ may be somewhat mitigated by a tuning of $\delta$.
By allowing the onset to occur more slowly, {\it i.e.}, with a larger value
of $\delta$, a slightly lower onset energy can be obtained. Nevertheless,
even with this tuning, the resulting onset energy, $E_{th}$, remains
significantly larger than in previous studies. Our value for the onset
energy might rather be considered as a lower bound, since, in principle,
the continuum could be pushed out to even higher energies and
compensated by adding further spectral strength to the excited states
({\it i.e.}, decrease $g_X$ or adding more states). However, a continuum
threshold at even higher energies is not well constrained by data.
Therefore we have taken the more conservative approach here.

\begin{table}[!tb]
\begin{center}
\begin{tabular}{|c|c|c|c|c|}
    \hline
    WSR &$0^{\rm th}$ & $1^{\rm st}$ & $2^{\rm nd}$ & $3^{\rm rd}$ \\
    \hline
    $\%$ agreement &-1.28\% & $\sim 0\%$ & $\sim 0\%$ & -96\% \\
    \hline
\end{tabular}
\end{center}
\caption{Percent disagreement between LHS and the RHS of the Weinberg-type
sum rules resulting from our fit.}
\label{tab:WSR}
\end{table}

\begin{figure}[tb]
\centering
\includegraphics[width=.45\textwidth]{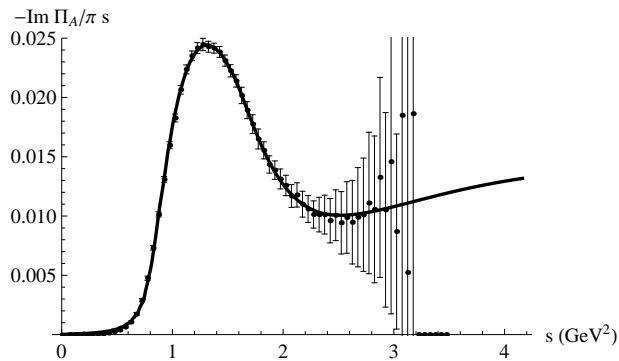}
\caption{Spectral function in the axial-vector channel without the
$a_1^\prime$, compared to experimental data~\cite{Barate:1998uf}.}
  \label{fig:a1spec2}
\end{figure}

Second, the current study is the first to include excited resonance states
into a construction of the spectral functions using sum rule techniques.
Previous works~\cite{Leupold:2001hj,Kwon:2008vq,Kwon:2010fw} have argued that
their continuum
ansatz, with a smaller threshold, was providing sufficient agreement
with the data such that explicitly treating these states was not necessary. However, when the continuum is pushed to higher energies,
 a region in energy is created where the vector spectral function is no
longer in good agreement with experiment ({\it cf.} Fig.~\ref{fig:rhospec}
for the continuum contribution undershooting the data). To accommodate
this, it is therefore natural to include the effects of a $\rho^\prime$
resonance. This argument, based upon \emph{observed} bumps in
experimental data, cannot be applied to the axial-vector channel, since
there is no direct indication which suggests a clear need for an
$a_1^\prime$ state {\it a priori}. However, our enforcing of the
Weinberg-type sum rules dictates additional strength in the
axial-vector channel which is not incompatible with data.
To further illustrate this point, let us switch off the $a_1^\prime$ peak.
The resulting axial-vector spectral function still describes the data as
seen in Fig.~\ref{fig:a1spec2}, but evaluating the Weinberg-type
sum rules in this scenario leads to appreciable discrepancies in all
cases, {\it cf.}~the dot-dashed curves in Fig.~\ref{fig:wsr}. Therefore, the
requirement of satisfying the sum rules by the constructed spectral
functions lets us deduce the presence of an $a_1^\prime$ state.

Third, one may ask if there is other evidence for an excited axial-vector state.
The Particle Data Group~\cite{pdg} lists three possible candidates,
$a_1 (1640)$, $a_1 (1930)$ and $a_1 (2095)$, though none of them are very
well established (the latter two can only be found in the ``Further States"
list and none are included in the summary table). Being the first excited
resonance, it would be natural to associate our $a_1^\prime$ with the
$a_1 (1640)$, but the mass and width do not match well. A similar problem
actually holds for the excited vector resonance: while we fit the ALEPH data
with a single $\rho^\prime$ state of mass 1565~MeV, the PDG has
states at $\rho (1450)$ and $\rho(1700)$. This suggests that the bump
seen in the data is an amalgamation of the two excited $\rho$
states. In principle, we could have tried to use  a spectral function
with two excited states, but the significance for the latter is not
apparent from the single bump in the $\tau$-decay data and thus it would
only serve to increase the number of our parameters without improving the fit.
Though we call our state $\rho^\prime$, it may well represent the
contribution from the two excited resonance states quoted by the PDG. A
similar argument holds for the $a_1^\prime$ state. It is then not a
problem that the properties of this proposed state do not match well with
previously seen individual resonances. Nevertheless, it is rather
consistent with an average of the lowest two states, similar to the
$\rho^\prime$ case.

Fourth, by adjusting the parameters of the $a_1^\prime$ states, the percent deviations of the Weinberg-type sum rules can change. The percent deviation of the second Weinberg-type sum rule is the most sensitive to such changes in parameters because the RHS of this sum rule is numerically very small. Each sum rule responds differently to each parameter. For example, the percent deviation of WSR-3 can be decreased by shifting the $a_1^\prime$ mass to lower energies, however the percent deviation of all the other sum rules becomes worse. As an extreme
case, the $a_1^\prime$ can be shifted to lower energies accompanied by the
reduction of the $a_1^\prime$ coupling so that WSR-2 and WSR-3 are both
``precisely" satisfied; WSR-0 and WSR-1 are then violated by $-2.1\%$ and
$-3.6\%$, respectively. Although the overall deviations are smaller in this
case than with the parameters in Table \ref{tab:para}, we believe that is
it better to get the best possible agreement with the two sum rules
which are most accurately know.

Finally, the other fitted parameters seem to be in reasonable agreement with
expectations. The $a_1$ mass is well within the PDG range~\cite{pdg}, while
its width is at the upper end of its range. The form-factor scales of
around 1 GeV are of typical hadronic size, with the one for the $a_1$
somewhat, though not unreasonably, smaller. The threshold scale is
consistent with decays into 4$\pi$ states for the $\rho^\prime$, while it
is slightly larger than the physically expected 5$\pi$ threshold for the
$a_1^\prime$.  Although the coupling for the $a_1^\prime$,
$g_{a_1^\prime}$, seems large compared to the other states, its numerical
value is dependent on the location of the continuum as discussed above.
Overall, the chosen parameters result in spectral functions that fit the
data and appear to be within reason of their physical interpretation.

%%%%%%%%%%%%%%%%%%%%%%%%%%%%%%%%%%%%%%%%
\section{QCD sum rules}
\label{sec:qcd}
%%%%%%%%%%%%%%%%%%%%%%%%%%%%%%%%%%%%%%%%%
Although the QCD sum rules were not used to initially constrain the model,
it is interesting to determine to what extent the resulting spectral
functions satisfy them. From Eqs.~(\ref{eq:rhoQCD}) and (\ref{eq:a1QCD}),
one can see that there are three condensates whose values are needed.
The quark condensate will be set at the commonly used value,
$\langle \bar{q} q \rangle = (-0.25 {\rm GeV})^3$.
However, there is appreciable variability in the values of $\kappa$ and the
gluon condensate in the literature.
%\cite{Klingl:1997kf,Hatsuda:1991ez,Leinweber:1995fn, Leupold:2001hj,Hilger:2010cn,Shifman:1978bx,Narison:1995jr,Narison:2011xe}.
Their values will be tuned to optimize the agreement of the QCD sum rules
given our input spectral functions.

To quantify this agreement, we will use the method described in
Refs.~\cite{Leinweber:1995fn,Leupold:1997dg}. In this method, one first
defines a LHS and a RHS corresponding to the resonance part in the dispersion
integral and the OPE with the continuum contribution from the dispersion
integral subtracted, respectively. For our purposes, this division will
be written as in Eqs.~(\ref{eq:rhoQCD}) and (\ref{eq:a1QCD}) with the
continuum part of the spectral functions subtracted from both sides.
Second, a range in the Borel mass, the so-called Borel window, needs to
be established where one expects the best agreement between the two sides
given their applicability limitations. We will define the lower limit,
$M_{\rm min}$, in the standard way, {\it i.e.}, as the Borel mass where the
terms proportional to $M^{-6}$ contribute at most 10\% to the RHS, indicating
that the OPE becomes unreliable at still smaller values of the Borel mass.
For large Borel mass, the OPE is dominated by the $M^0$ term. However,
this term is nearly trivially satisfied by the continuum. Therefore, in
order to analyze the agreement due to the resonance states, an upper
limit to the Borel window is introduced. This is typically chosen as the
mass where the continuum contribution equals the one from the resonances.
We have found that applying this procedure with our spectral functions
results in a Borel window nearly double in size of what was previously
found in the literature. This difference simply arises due to the significantly higher
energies for the onset of our universal continuum. We therefore decided
to redefine the upper limit of the window, $M_{\rm max}$, as the mass
where the continuum contribution is half of the resonance contribution.
This produces a Borel window of similar size as in previous studies.
Instead of a 50/50 split between continuum and resonances at the upper
range, we thus have a 67/33 proportion between resonances and continuum.
With these considerations, we found a Borel window of
$0.85~{\rm GeV} < M < 1.47~{\rm GeV}$ for the vector channel and
$0.89~{\rm GeV} < M < 1.48~{\rm GeV}$ for the axial-vector channel,
which is comparable to earlier
studies~\cite{Leinweber:1995fn,Leupold:1997dg,Leupold:2001hj}.
\begin{figure}[!tb]
\centering \subfigure[Vector Channel]{
\includegraphics[width=.45\textwidth]{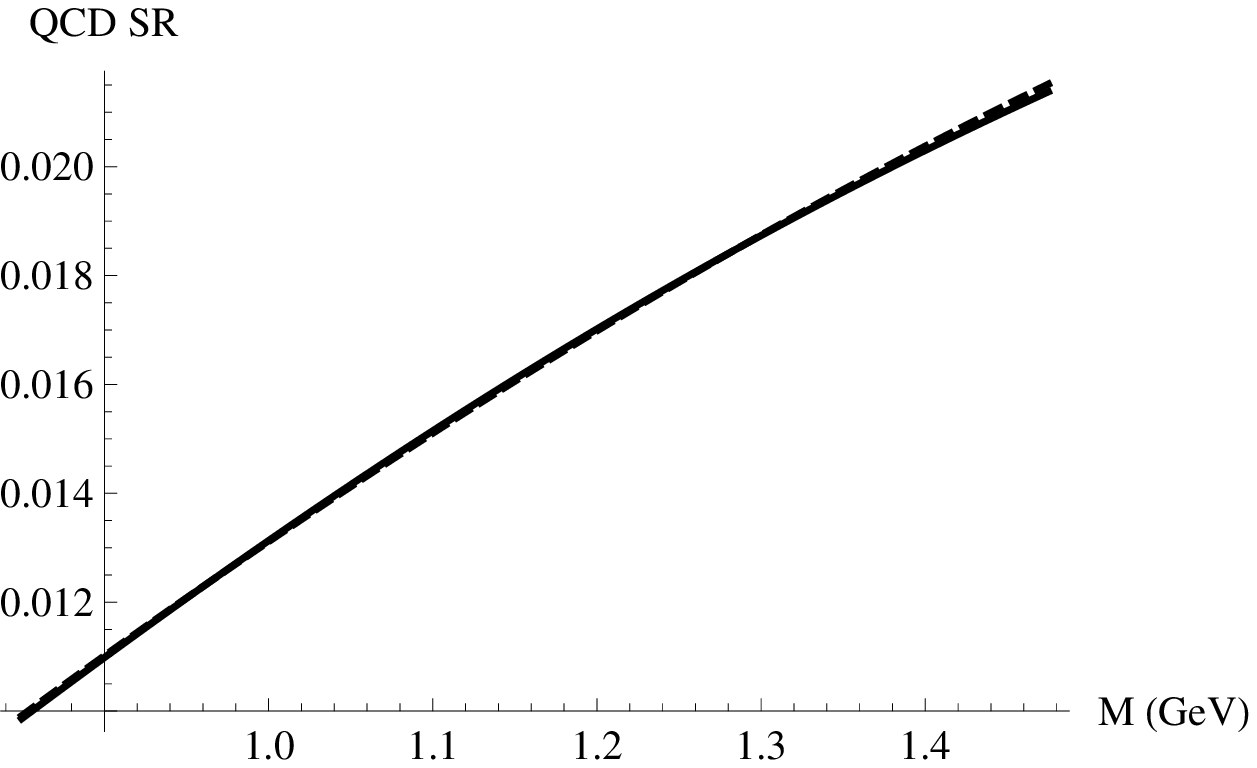}}
\subfigure[Axial-Vector Channel]{
\includegraphics[width=.45\textwidth]{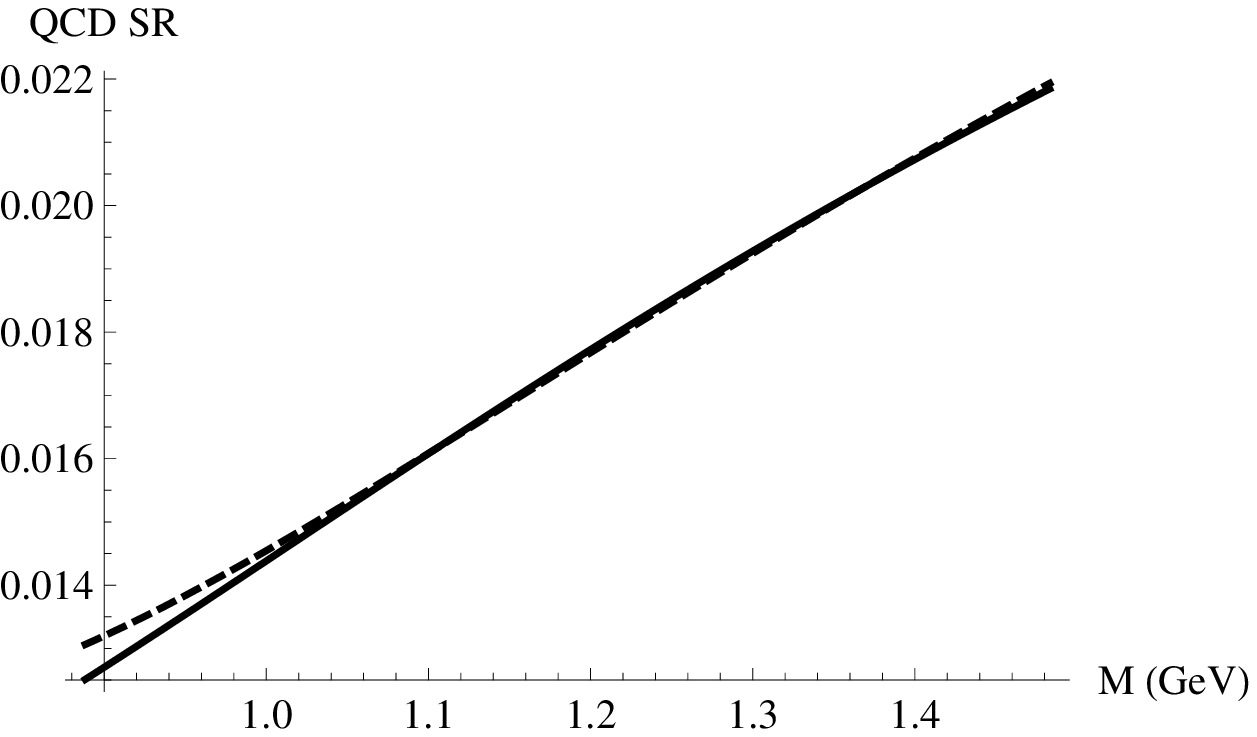}}
\caption{Graphical representation of the QCD sum rules as a function of
Borel mass over the Borel window. The LHS (dispersion integral using the
spectral functions of the previous section) of the sum rule is given by
the solid curve while the RHS (OPE) is represented by the dashed curve.}
  \label{fig:qcd}
\end{figure}

With spectral functions, condensate values, and the Borel window defined, the
agreement between the two sides of the sum rule is measured by the
value $d$ defined as
\begin{equation}
d = \frac{1}{\Delta M^2} \int^{M_{\rm max}^2}_{M_{\rm min}^2}
dM^2 |1- {\rm LHS}/{\rm RHS}| \ ,
\end{equation}
where $\Delta M^2 = M_{\rm max}^2 - M_{\rm min}^2$. One can think of
$d$ as the average deviation between the two sides over the Borel window.
The optimization of the values for $\kappa$ and the gluon condensate
thus amounts to minimizing $d$. Optimizing the vector and axial-vector
channels independently yields different values of $\kappa$ and the gluon
condensate for the two channels. While the gluon condensate should be identical,
the value of $\kappa$, representing correlations beyond the
ground state, can, in principle, be different for different quantum
numbers. For simplicity, we have decided to also assume a universal
$\kappa$ value and to minimize the sum of the $d$ values obtained from the
two channels, that is, we minimize $d_{\rm tot} = d_V + d_A$. To estimate
the uncertainty
in the optimized values, the range in $\kappa$ and the gluon condensate which
produces a combined deviation of less than $1\%$ has been found.
The minimization procedure gives $\kappa = 2.1^{+0.3}_{-0.2}$ and
$\langle \frac{\alpha_s}{\pi} G^2_{\mu\nu} \rangle
= 0.022\pm 0.002\, {\rm GeV}^4$.
The optimized $d$ values for vector and axial-vector channel are 0.24\% and
0.56\%, respectively. Increasing the upper limit of the Borel window, $M_{\rm max}$ by
$5\%$ yields no significant change in the values of the parameters, but
increases the deviation. Furthermore, when $\kappa_V$ and $\kappa_A$ are
decoupled, the optimization procedure yields rather similar results,
$\langle \frac{\alpha_s}{\pi} G^2_{\mu\nu} \rangle = 0.022\, {\rm GeV}^4$,
$\kappa_V = 2.1$, $\kappa_A = 2.0$, $d_V = 0.24\%$, and $d_A = 0.55\%$,
thereby justifying {\it a posteriori} the simplifying assumption of identical
$\kappa$ values.

Our values for $\kappa$ are well within the range previously found in the
literature~\cite{Klingl:1997kf,Hatsuda:1991ez,Leinweber:1995fn,Leupold:2001hj},
while the value for the gluon condensate is in the upper range. For example,
early applications in charmonium sum rules have extracted
$\langle \frac{\alpha_s}{\pi} G^2_{\mu\nu} \rangle \simeq
0.012\, {\rm GeV}^4$~\cite{Shifman:1978bx}. On the other hand, more recent
studies of higher moments of the charmonium sum rules yield significantly larger
values, of around
$0.022 \,{\rm GeV}^4$~\cite{Narison:1995tw,Narison:2011xe,Narison:2011rn},
close to our value.

To illustrate the agreement of the QCD sum rules, we display in Fig.~\ref{fig:qcd}
both LHS and RHS for each channel, scaled by $M^2$, as a function of $M$
over the Borel window. With the sub-$1\%$ agreement, it is difficult to
distinguish the two curves in each figure. We note that an increase in $\kappa$
with fixed gluon condensate will shift the
RHS of the vector channel down, while the RHS of the axial channel is shifted
up (dashed curves),
thereby making the agreement in {\emph both} channels worse. Overall, the spectral
functions constructed may be considered consistent with the QCD sum rules.

%%%%%%%%%%%%%%%%%%%%%%%%%%%%%%%%%%%%%%%%%
\section{Conclusion}
\label{sec:conc}
%%%%%%%%%%%%%%%%%%%%%%%%%%%%%%%%%%%%%%%%%
In the present work, we have performed a combined analysis of Weinberg-type
and QCD sum rules using vacuum spectral functions for vector and axial-vector
channels constructed via quantitative fits to hadronic $\tau$-decay data. For
the ground-state resonances, we employed a microscopic $\rho$ spectral function
and a Breit-Wigner ansatz for the $a_1$. A novel feature of our approach is the
introduction of an excited state in each channel which, in turn, allowed us to
employ an universal perturbative continuum part at high energies. The
universality of the continua is rather welcome in view of chiral
degeneracy in perturbation theory, and leads to a higher onset energy than in
previous works. While the excited vector state ($\rho^\prime$) was deduced from
an accurate fit to data, we have found that the three lowest Weinberg-type sum
rules can be quantitatively satisfied only if the existence of an $a_1^\prime$
state is postulated. The latter lies outside the direct realm of the
axial-vector $\tau$-decay data, but the extracted mass is compatible with (and
in a sense confirms) an average of previously
observed states. Furthermore, the resulting spectral functions were implemented
into QCD sum rules, resulting in a sub-$1\%$ agreement with the operator product
expansion (comparable to other state-of-the-art analyses). We believe that these
spectral functions provide a good basis for future studies of medium
modifications, to shed light on the long-standing problem of testing
chiral symmetry restoration with dilepton data.

\acknowledgments
This work is supported by the US-NSF under grant No.~PHY-0969394 and
by the A.-v.-Humboldt Foundation (Germany).

\end{document}